# High-power diode-pumped Er$^{3+}$:YAG single-crystal fiber laser

Igor Martial,[1,2,*] Julien Didierjean,[2] Nicolas Aubry,[2] François Balembois,[1] and Patrick Georges[1]

[1]Laboratoire Charles Fabry de l'Institut d'Optique, CNRS, Université Paris-Sud, Campus Polytechnique, RD 128, 91127 Palaiseau Cedex, France

[2]Fibercryst SAS, La Doua-Bâtiment l'Atrium, Boulevard Latarjet, F- 69616 Villeurbanne Cedex, France

## ABSTRACT

We describe an efficient laser emission from a directly grown Er$^{3+}$:YAG single-crystal fiber that is resonantly pumped using a continuous-wave (CW) laser diode at 1532 nm. In a longitudinal pumping, it emits 12.5 W at 1645 nm with a slope efficiency of 32%, which is the highest ever reported for a directly grown Er:YAG single-crystal fiber laser. Using an off-axis pumping scheme, CW output powers up to 7.3 W can be reached and in Q-switched operation, the laser produces 2 mJ pulses with a duration of 38 ns at the repetition rate of 1 kHz with an M² factor below 1.8. To our knowledge this is the first directly grown Er$^{3+}$:YAG single-crystal fiber Q-switched laser. In dual-side pumping scheme a laser emission at 1617 nm is achieved with output powers up to 5.7 W representing the highest output power ever achieved by a diode-pumped Er:YAG laser at this wavelength.

## 1. INTRODUCTION

The yttrium aluminium garnet (YAG) crystal doped with Er$^{3+}$ is one of the most attractive materials for the development of multi-watt compact eye safe laser sources around 1.6 μm. Such sources are required for many applications like remote sensing, free-space communications, range detection, and designation. The optical transition between the $^4I_{15/2}$–$^4I_{13/2}$ manifolds includes the possibility of in-band pumping (between 1.45 and 1.53 μm). This minimizes the heat generated by the pumping process and hence limits the thermally induced birefringence and the thermal lens. Er:YAG lasers can be resonantly pumped using erbium-doped fiber lasers (EDFL) [1-5]. The excellent beam quality of EDFL allows good spatial overlap of the pumped volume with the laser mode and efficient pump absorption due to the narrow bandwidth. High power and good optical efficiency have been achieved using this EDFL pumping [2]. However, this pumping approach adds complexity, weight and volume to the laser system. Additionally, the overall optical efficiency with EDFL pumping is typically only about 28% due to the number of stages [4].

Direct pumping using high power laser diodes can be a solution to improve the overall efficiency of the system [6-9]. However, such direct pumping is complicated because of the low brightness of laser diodes. For this purpose, the use of long and thin rods (single-crystal fibers), confining the pump beam close to the signal beam, represents one way to manage the poor beam quality of pump laser diodes and to achieve high pump intensities along the whole laser medium [10]. However, these rods are difficult to manufacture from standard Czochralski-grown boules: after cutting and complex polishing of the cylinder, it is difficult to reduce the diameter below 1 mm for a few centimeter long sample. The micro-pulling-down technique (μPD) [11] provides a solution to this problem as laser rods can be elaborated directly, without any additional polishing of the cylinder. The diameter can be easily reduced to a few hundreds of microns and typical crystal lengths can be up to 1 meter. In a previous work [12], we demonstrated the growth of laser quality Er:YAG single-crystal fiber by the micro-pulling-down technique and the first laser ever achieved with such a gain medium.

In this paper, we improve the output power in CW by more than one order of magnitude. We report Q-switched operation of a directly grown Er:YAG single-crystal fiber for the first time to our knowledge. Two configurations are tested: with an on-axis pumping scheme and with an off-axis pumping scheme. We report laser emission not only at 1645 nm but also at 1617 nm corresponding to a better atmospheric transmission.

## 2. ON-AXIS PUMPING SCHEME

### 2.1 Experimental setup

The experimental setup is shown in Fig. 1. It consists of an end-pumped actively cooled 0.5% $Er^{3+}$:YAG single-crystal fiber which has a diameter of 800 μm, a length of 60 mm and anti-reflection coatings on both ends. The fiber was produced by Fibercryst by the micro-pulling-down technique. We use a bi-concave cavity folded with a dichroic mirror through which we pump the single-crystal fiber. The first concave mirror is highly reflective at 1645 nm and the second one has a transmission of 20%. The pump light is provided by a fiber-coupled laser diode, with a 400 μm core diameter and a numerical aperture of NA = 0.22, delivering up to 80 W at 1532 nm (~1 nm FWHM). The fiber is imaged into the crystal by a 1:1 telescopes consisting of two AR coated lenses L1 and L2 with effective focal lengths of 50 mm. In order to maximize the output power, the focus point is put slightly inside the single-crystal fiber, close to the pumped face. It corresponds to a 400 μm diameter beam. After free space propagation in the first 10 mm of the single-crystal fiber, the pump beam is guided by total internal reflections inside the gain medium.

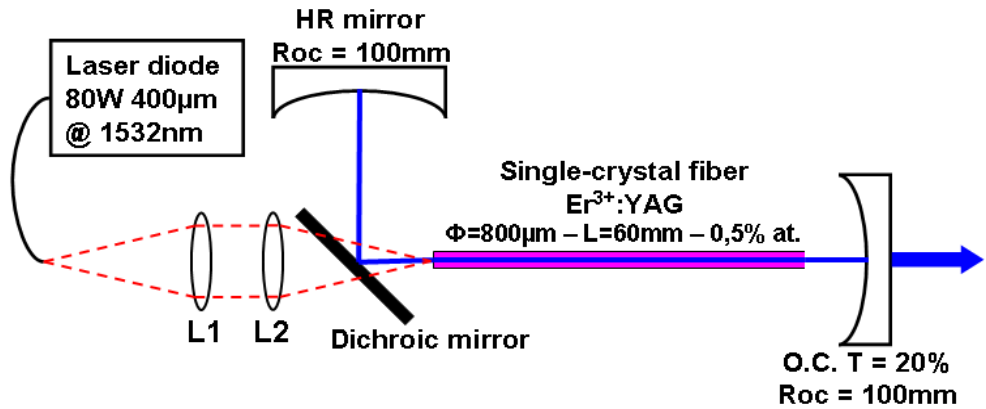

Fig. 1: Experimental setup in on-axis pumping configuration.

### 2.2 Results

A continuous wave laser operation occurs at the wavelength of 1645 nm. Fig. 2 reports the measured output power versus the absorbed pump power. The laser threshold is measured at 6.3 W and the maximum output power at 1645 nm is 12.5 W for 45.8 W of absorbed pump power, leading to a slope efficiency of 32%. To our knowledge, this is the highest power in laser operation ever achieved with directly grown Er:YAG single-crystal fibers.

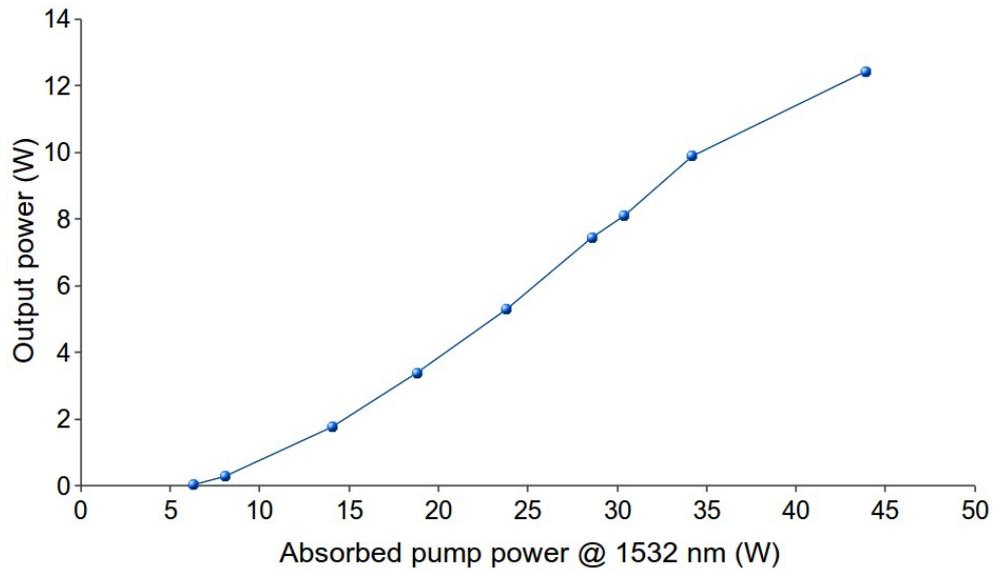

Fig. 2: Output power at 1645 nm versus absorbed pump power in CW regime in the on-axis configuration.

In order to realize Q-switched operation, an acousto-optic modulator (A.O.M.) is placed inside the cavity. The A.O.M. brings 7% additional simple pass losses to the cavity. In this configuration, in CW operation the maximum output power at 1645 nm is 7.3 W for 45 W of absorbed pump power. In Q-switched regime, an average power of 7.3 W is reached whatever the repetition rate between 10 kHz and 80 kHz. The dependence of average output power and single pulse energy versus the repetition rate is shown on Fig. 3. For the maximum absorbed pump power (45 W), the laser could not be operated at lower repetition rates than 10 kHz due to damage of the optical coating on the dichroic mirror caused by the intra-cavity peak power of the pulses. At 10 kHz, the pulse energy is 800 µJ.

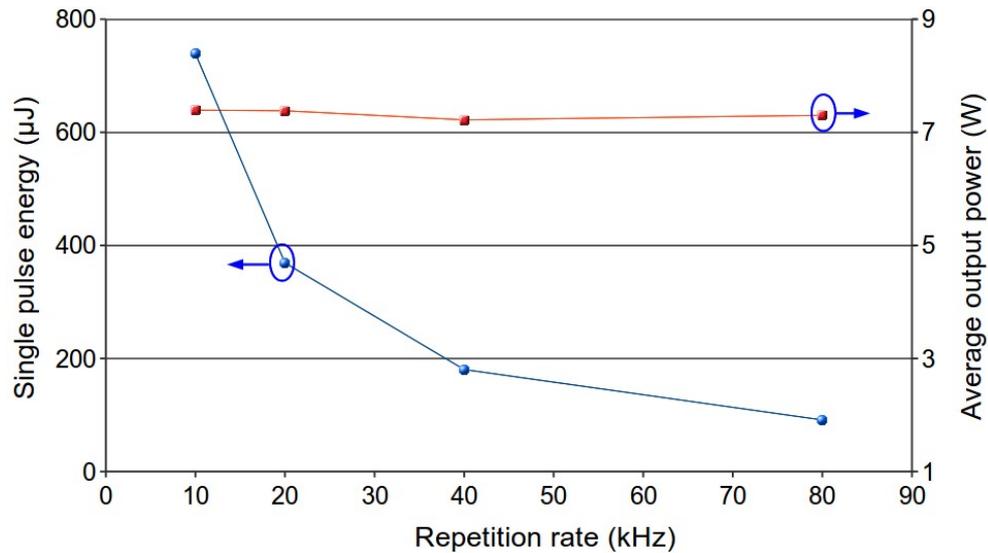

Fig. 3: Average output power (red square, right axis) and output single pulse energy (blue circle, left axis) at 1645 nm versus the repetition rate in Q-switched regime and in on-axis pumping. The absorbed pump power is 45 W and the output coupler transmission is 20%.

# 3. OFF-AXIS PUMPING SCHEME

## 3.1 Experimental setup

The manufacturing of the dichroic mirror is complex because of the small difference between the pump wavelength and the signal wavelength and leads to coatings with low damage threshold. As we demonstrated before, this is a limitation in Q-switched regime where the intra-cavity intensity is very high. Single-crystal fibers provide a solution to this problem by combining the guiding of the pump beam and a very large numerical aperture. This allows off-axis pumping of the gain medium without any dichroic mirror. The experimental setup is shown in Fig. 4. The setup is exactly the same as presented before except for the dichroic mirror.

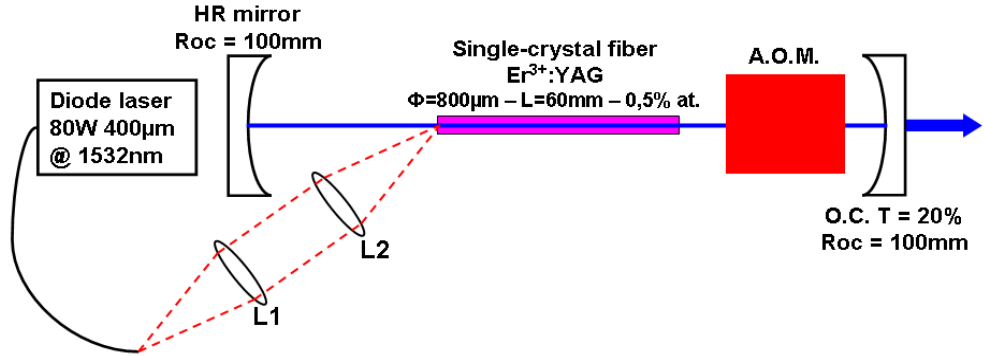

Fig. 4: Experimental setup in off-axis pumping configuration.

## 3.2 Results

In this configuration, in continuous wave laser operation the laser threshold is measured at 15.8 W and the maximum output power at 1645 nm is 7.3 W for an absorbed pump power of 58 W. This leads to a slope efficiency of 18% (Fig. 5). This lower efficiency, compared to the longitudinal pumping scheme, is due to the overlap between the pump and the signal beam which is better in longitudinal pumping than in off-axis pumping.

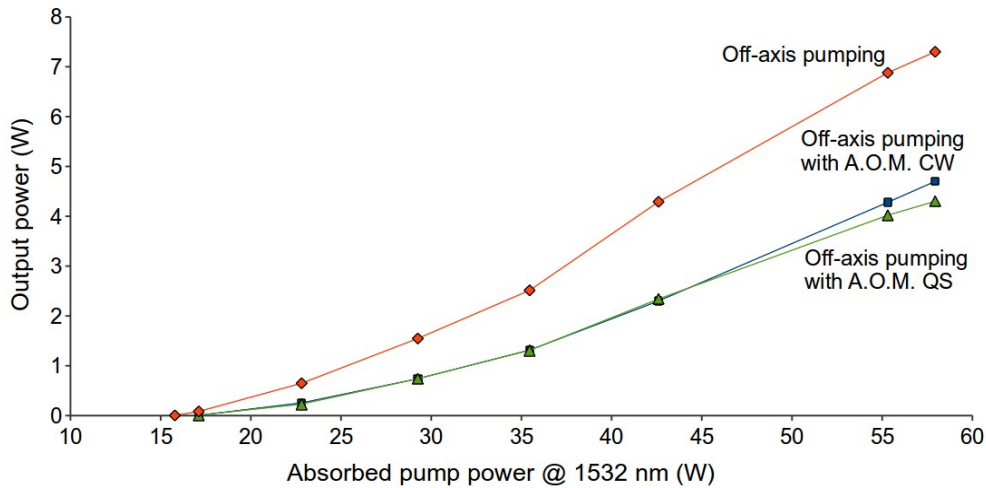

Fig. 5: Output power at 1645 nm versus absorbed pump power at 1532 nm in off-axis pumping without A.O.M. (red diamond-shaped), in off-axis pumping with A.O.M. (blue square) and in off-axis pumping Q-switched regime at 5 kHz (green triangle). The output coupler transmission is 30%.

In order to realize Q-switched operation in the off-axis pumping configuration, the A.O.M. is placed inside the cavity and an output coupler with a higher transmission than in the previous experiment (T = 30% @ 1645 nm) is used to prevent any damage on optics and coatings. In this configuration, in CW operation the measured laser threshold is 17 W and the maximum output power at 1645 nm is 4.7 W for 58 W of absorbed pump power (see Fig. 5). In Q-switched regime, at the repetition rate of 5 kHz, the output average power is nearly equal to the output power in CW regime. This indicates that there is no significant losses due to up-conversion or amplified stimulated emission (ASE) when Q-switching at 5 kHz. The dependence of average output power and single pulse energy versus the repetition rate is shown on Fig. 6. For the maximum absorbed pump power (58 W), the laser could not be operated at lower repetition rates than 1 kHz due to damage of the optical coatings of the single-crystal fiber caused by the intra-cavity peak power of the pulses.

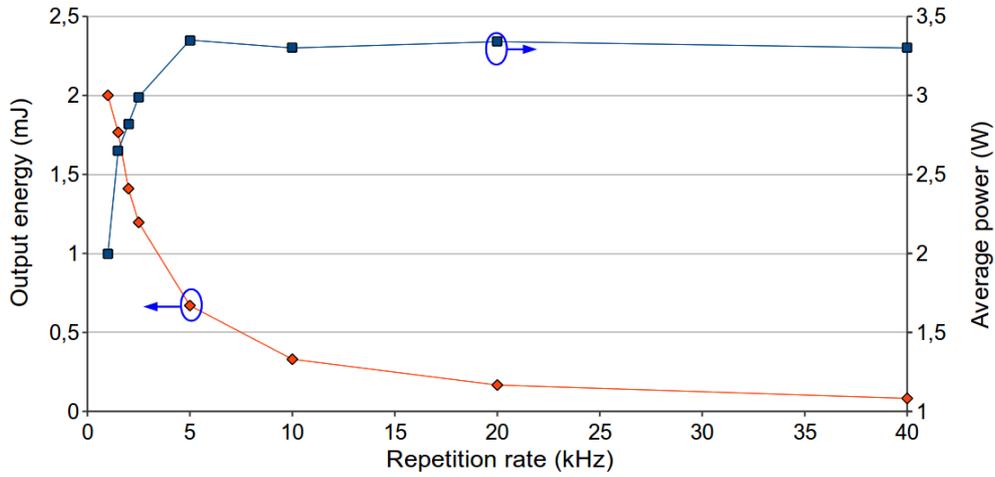

Fig. 6: Average output power (blue square, right axis) and output single pulse energy (red diamond-shaped, left axis) at 1645 nm versus the repetition rate in Q-switched regime and in off-axis pumping. The absorbed pump power is 58 W and the output coupler transmission is 30%.

At 1 kHz, the pulse energy is 2 mJ with a pulse duration of 38 ns. Fig. 7 shows both spatial and temporal profile of a 2 mJ pulse obtained at 1 kHz. In this configuration, the pulse duration is similar to the one obtained with efficient EDFL pumping system [4] and shorter than in recent diode pumping system [10,13] and leads to peak-power up to 53 kW. The beam presented a good spatial quality with an M² factor lower than 1.8 in both directions.

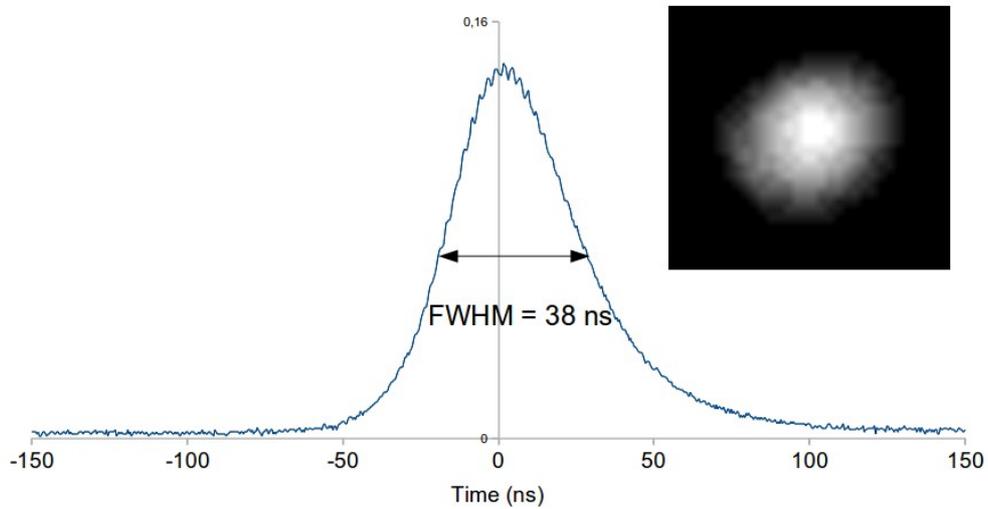

Fig. 7: Temporal profile of a 2 mJ pulse obtained at 1 kHz under 58 W of absorbed pump power. Inset : spatial profile of the output beam at 1645 nm.

## 4. LASER EMISSION AT 1617 NM

### 4.1 Experimental setup

Imaging system based on Er:YAG lasers operating at 1645 nm are limited by some atmospheric absorption lines due to methane [14]. The typical atmospheric absorption at 1645 nm is 0.1 km$^{-1}$. Fortunately, Er:YAG has also a laser transition at 1617 nm which lies in a region of the spectrum where there is no atmospheric absorption line. This transition benefits from a higher emission cross-section, but its lower level is more populated. It requires ~14% of the Er$^{3+}$ ions to be excited to the upper manifold to reached transparency compared with ~9% for the 1645 nm [15]. In order to increase the inversion population inside the gain medium to obtain laser emission at 1617 nm, we use thiner single-crystal fiber. By reducing the diameter of the single-crystal fiber, the pump intensity is increased. Moreover a dual-side pumping scheme of the single-crystal fiber is used in order to reduce the thermal load. The experimental setup is shown in Fig. 8. The pump beam is divided in two parts by a mirror highly reflective at 1532 nm at 45°. Compared to a partially reflective beam splitter, this setup allows to keep the same brightness for the two separated beams with respect to the incident beam. We use a bi-concave cavity with two plane dichroic mirrors for both side pumping. We test different output couplers as reported on the Fig. 9. In order to select the 1617 nm transition, we use a 50 μm thick uncoated etalon in fused silica.

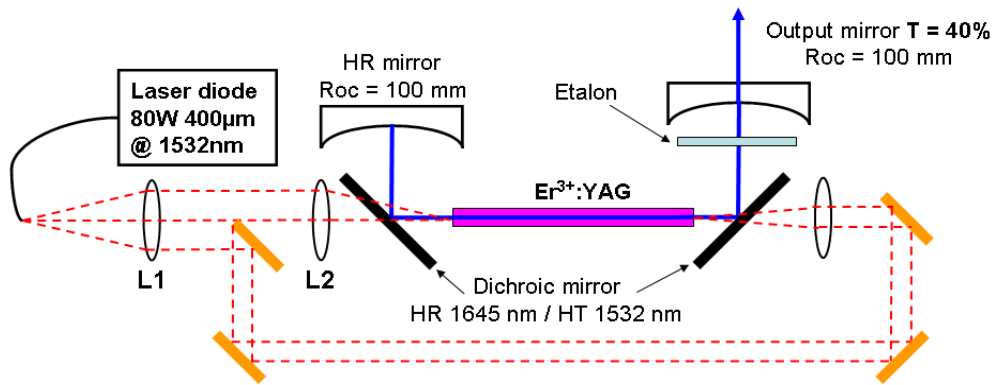

Fig. 8: Experimental setup in dual-side pumping scheme for laser emission at 1617 nm.

### 4.2 Results

In this configuration we tested 6 cm long, 0.5% at. single-crystal fiber with diameter of 800 and 600 μm. We first tested different output coupler transmissions (see Fig. 9) at the operating wavelength of 1645 nm (with etalon). The optimum output coupler transmission is 20% for 800 μm diameter single-crystal fiber and 40% for the 600 μm diameter. This shows that the gain is clearly higher with thinner single-crystal fiber.

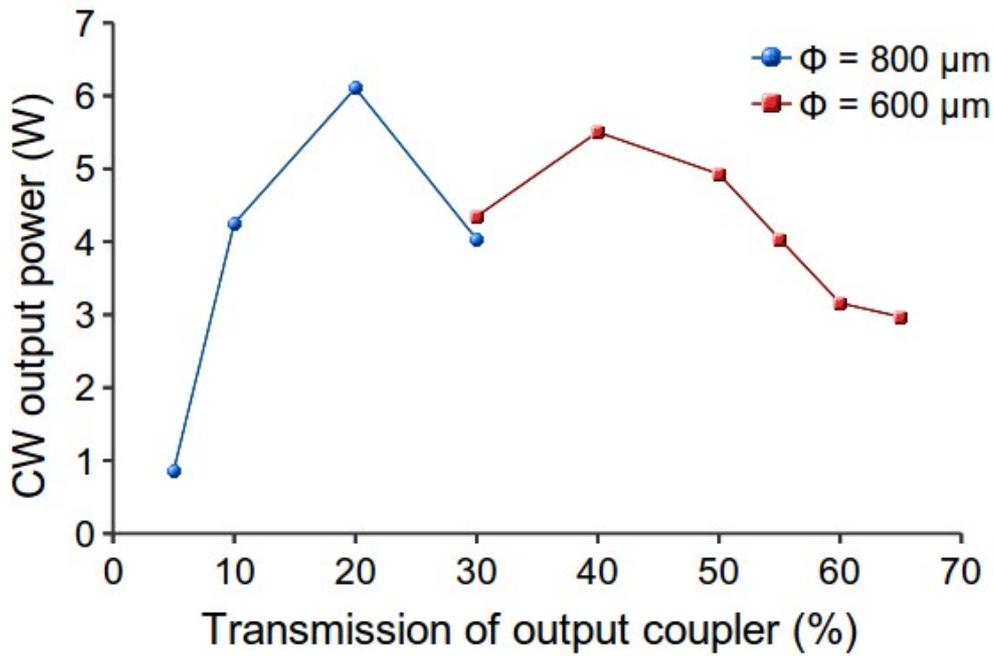

Fig. 9: CW output power versus the transmission of the output coupler in the dual-side pumping scheme at the wavelength of 1645 nm with 800 μm diameter single-crystal fiber (blue circle) and 600 μm diameter single-crystal fiber (red squares).

Thanks to this high gain, laser emission at 1617 nm is possible. Thanks to an appropriate orientation of the etalon, the 1627 nm emission can be selected. The Fig. 10 reports the measured output power versus the absorbed pump power in continuous wave regime at this wavelength. The laser threshold is measured at 15 W of absorbed pump power and the maximum output power is 5.7 W under 56 W of absorbed pump power leading to a slope efficiency of 15.4%. To our knowledge, this is the highest power in laser operation ever achieved with diode pumped Er:YAG laser at 1617 nm.

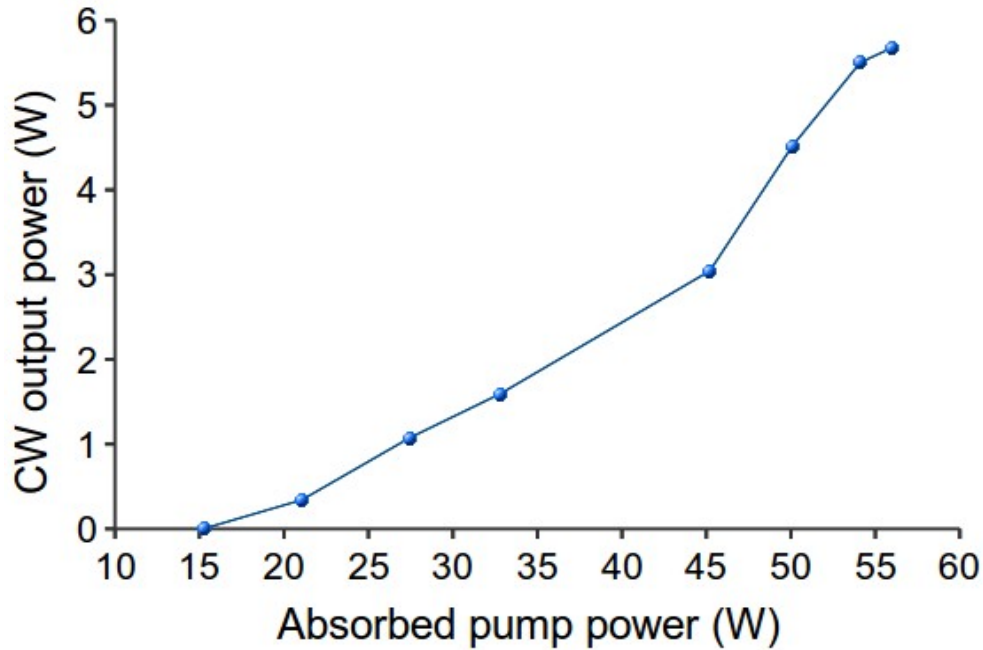

Fig. 10: Output power at 1617 nm versus absorbed pump power in continuous wave regime in the dual-side pumping scheme with the 50 μm thick etalon and an output coupler transmission of 40%.

## 5. CONCLUSION

In conclusion, we have demonstrated an efficient Er:YAG single-crystal fiber laser that is resonantly pumped using a CW laser diode at 1532 nm. In a longitudinal pumping configuration, CW output powers up to 12.5 W at 1645 nm have been obtained with a slope efficiency of 32%. The overall optical efficiency is 25 %, close to the one obtained with EDFL pumping but in a more simple setup. Efficient Q-switched operation has been realized, but is limited by the damage threshold of the dichroic mirror. Using an off-axis pumping scheme CW output powers up to 7.3 W have been obtained. In pulsed operation, the laser has been Q-switched at the repetition rate of 5 kHz with an average power of 3.4 W. The highest single pulse energy obtained was 2 mJ with a pulse duration of 38 ns at the repetition rate of 1 kHz and with a $M^2$ factor below 1.8 in both directions. To our knowledge, this is the first Q-switched directly grown Er:YAG single-crystal fiber laser. Higher pulse energy should be achieved by employing a more-transparent output coupler. In dual-side pumping scheme with thinner single-crystal fiber CW laser emission was obtained at 1617 nm with output powers up to 5.7W.

## 6. ACKNOWLEDGEMENT

The authors are grateful to M. Eichhorn from the French-German Research Institute of Saint-Louis (ISL) (Saint-Louis, France) for the loan of the dichroic mirror.

This work is supported by the DGA under the project FEYPIA n° 2008 34 0019.